\documentclass[12pt,a4paper]{article}

\usepackage{amsbsy,latexsym,cite}

\newcommand{\no}{\noindent}

\newcommand{\mod}[1]{\vert {#1}\vert}
\newcommand{\pa}{\partial}

\newcommand{\ee}{\mathrm{e}}

\newcommand{\ket}[1]{\vert #1 \rangle}
\newcommand{\bra}[1]{\langle #1 \vert}

\newcommand{\xb}{{\boldsymbol x}}

\newcommand{\pb}{{\boldsymbol p}}
\newcommand{\qb}{{\boldsymbol q}}

\newcommand{\kb}{{\boldsymbol k}}

\newcommand{\nub}{{\boldsymbol \nu}}

\newcommand{\intp}{\int\!\frac{\mathrm{d}^dp}{(2\pi)^d}\;}
\newcommand{\intdq}{\int\!\frac{\mathrm{d}^dq}{(2\pi)^d}\;}

\begin{document}

\begin{titlepage}


\vskip19truemm
\begin{center}{\Large{\textbf{Anti-Screening by Quarks and the Structure of
the\\[2mm]
Inter-Quark Potential}}}\\ [12truemm] \textsc{Emili
Bagan,$^a$}\footnote{email: bagan@ifae.es} \textsc{Martin
Lavelle$^b$}\footnote{email: mlavelle@plymouth.ac.uk}
and \textsc{David McMullan$^b$}\footnote{email: dmcmullan@plymouth.ac.uk}\\[5truemm]
\textit{$^a$Grup de F\'\i sica Te\`orica and IFAE\\Edificio Cn\\
Universitat Aut\`onoma de Barcelona\\ E-08193 Bellaterra (Barcelona)\\ Spain}\\[3truemm] \textit{$^b$Particle Theory Group\\
School of
Mathematics and Statistics\\ The University of Plymouth\\
Plymouth, PL4 8AA\\ UK} \end{center}

\bigskip\bigskip\bigskip
\begin{quote}
\textbf{Abstract:} The inter-quark potential is dominated by
anti-screening effects which underly asymptotic freedom. We
calculate the order $g^6$ anti-screening contribution from light
fermions and demonstrate that these effects introduce a non-local
divergence. These divergences are shown to make it impossible to
define a coupling renormalisation scheme that renormalises this
minimal, anti-screening potential. Hence the beta function cannot be
divided into screening and anti-screening parts beyond lowest order.
However, we then demonstrate that renormalisation can be carried out
in terms of the anti-screening potential.
\end{quote}

\end{titlepage}

\bigskip\bigskip\medskip

\noindent Asymptotic freedom is the paradigm effect of QCD. It has
been shown in many approaches
\cite{Gribov:1977mi,Hughes:1980ms,Nielsen:1981sx,Drell:1981gu,%
Vainshtein:1982wh,Brown:1997gm,Zarembo:1998bp,Mansfield:1998vc,%
Lavelle:1998dv,Bagan:2000nc,Bagan:2001wj,Arnone:2002cs}
that the leading order beta function can be divided into screening
and anti-screening effects. In QCD with $n_f$ light fermions this
decomposition reads:
\begin{equation}\label{betasplit}
\beta(g)=-\frac{g^3}{(4\pi)^2}\left[4-\frac13-\frac{2n_f}3
\right]\,,
\end{equation}
where the dominant term (the $4$) corresponds to anti-screening and
the smaller $-\frac13-\frac{2n_f}3$ terms describe screening by glue
and by matter. It is well known that the gluonic screening effects
are due to physical (gauge invariant) glue.
Anti-screening~\cite{Lavelle:1998dv} is due to the contribution of
glue which is needed to construct a gauge invariant definition of a
coloured charge~\cite{Lavelle:1997ty}.

The static inter-quark potential~\cite{Peter:1997ig,Schroder:1998vy}
would seem to offer a direct way to study the screening and
anti-screening effects in QCD. The potential can be calculated via
Wilson loops. At order $g^4$ the momentum space bare potential,
$\tilde V^0$, in $D=4-2\epsilon$ dimensions is, up to some finite
terms,
\begin{equation}\label{bareV}
\tilde V^0(\kb^2)=-4\pi C_F\frac{\alpha_0}{\kb^2}\left\{
1-\frac{\alpha_0}{\pi}\left(\frac{11}{12}C_A-
\frac{n_f}6\right)\left[\log\left(\frac{\kb^2}{\nu^2}\right)
-\frac1{\epsilon}\right] \right\}\,,
\end{equation}
where  $\nu$ is a dimensional scale parameter and, in $SU(N)$,
$C_F=(N^2-1)/(2N)$ and $C_A=N$. This may be renormalised by the
standard charge renormalisation where the bare coupling
$\alpha_0=\nu^{-2\epsilon}Z_\alpha\alpha$ where
\begin{equation}\label{Zfull}
Z_\alpha=
1-\frac{\alpha}{\pi}\left(\frac{11}{12}C_A-\frac{n_f}{6}\right)
\frac1{\epsilon}\,.
\end{equation}
This yields to order $\alpha^2$
\begin{equation}\label{Vfinite}
\tilde V(\kb^2) =-4\pi C_F\frac{\alpha}{q^2} \left\{
1-\frac{\alpha}{\pi}\left(\frac{11}{12}C_A-\frac{n_f}6\right)\log\left(\frac{q^2}{\nu^2}\right)
\right\}\,.
\end{equation}
The Wilson loop approach does not, however, display the
screening/anti-screening decomposition of the potential. We will
therefore study here the interaction between two gauge invariant
descriptions of the colour charges. This has previously been
seen~\cite{Lavelle:1998dv,Bagan:2001wj} to show the decomposition of
screening and anti-screening in $3+1$ dimensions as well as in the
potential in $2+1$ dimensions~\cite{Bagan:2000nc} (where the beta
function vanishes).

The structure of this paper is as follows. We first review how a
gauge invariant description of physical charges directly shows how
anti-screening effects are produced. Then we calculate, for the
first time, how at order $g^6$ anti-screening effects due to light
fermions arise in the quark potential and that, due to non-local
divergences, a minimal (or anti-screening) charge renormalisation
approach breaks down here. Renormalising the potential directly is
shown to consistently handle the non-local structures and the full
result for the renormalised potential at this order is given.

It has previously been shown~\cite{Bagan:1999jf,Bagan:1999jk} that
the correct gauge invariant description of static charges in the
ground state is given by $h^{-1}\psi$ where $h^{-1}$ is a field
dependent dressing that surrounds the matter field $\psi$. The
dressing has a rich structure which in QED is as follows:
\begin{equation}\label{hqed}
  h^{-1}(x)=\exp\left(i\ee\int_{-\infty}^{x^0}\!ds
  \frac{\pa^iF_{i0}}{\nabla^2}(s,\xb)\right)
  \exp\left(-i\ee\frac{\pa_i A_i}{\nabla^2}\right)\,.
\end{equation}
Here the second exponential is the \emph{minimal dressing} which
ensures that the minimally dressed matter field,
$\exp\left(-i\ee\frac{\pa_i A_i}{\nabla^2}\right) \psi$, is gauge
invariant. The other factor, which we call the \emph{additional
dressing}, is itself gauge invariant.\footnote{The dressing may be
obtained from the requirement of gauge invariance plus an additional
dressing equation which may be derived~\cite{Bagan:1999jf} from the
heavy charge effective theory or from a study~\cite{Horan:1999ba} of
the asymptotic dynamics of charged particles.}

The structure of the dressing is reflected in calculations of the
potential between charges: anti-screening coming from the minimal
dressing~\cite{Lavelle:1998dv} while the effects of screening were
shown to come from the additional gauge invariant
dressing~\cite{Bagan:2001wj}. The different factors in the dressing
also make their presence separately felt in the infra-red structure
of the on-shell Green's functions of dressed fields: soft
divergences are cancelled by the effects of the minimal dressing,
while phase divergences are removed by the additional dressing.

The description (\ref{hqed}) has been extended to QCD order by
order in perturbation theory. With the inclusion of colour, the
\emph{minimal dressing} up to order $g^3$ is given
by~\cite{Lavelle:1997ty}
\begin{equation}\label{gthree}
h^{-1}(x)=  \exp\left( g \chi(x) \right)+O(g^4)\,,
\end{equation}
with $\chi=\chi^a T^a=(\chi_1^a+g\chi^a_2+g^2\chi^a_3)T^a$ and
\begin{equation}\label{chibits}
  \chi_1^a=\frac{\partial_j A_j^a}{\nabla^2}
  \,,\quad \chi_2^a=f^{abc}\chi^{bc}\quad
\mathrm{and} \quad \chi_3^a=f^{acb}f^{cef} \chi^{efb}
  \,,
\end{equation}
where we have defined
\begin{equation}\label{chitwo}
\chi^{bc}=\frac{\partial_j}{\nabla^2}\left(
\chi_1^bA_j^c+\frac12(\partial_j\chi_1^b)\chi^c_1 \right)\,,
\end{equation}
and
\begin{eqnarray}\label{chithree}
\chi^{efb}&=&\frac{\partial_j}{\nabla^2} \Big( \Big.
\chi^{ef}A_j^b +\frac12 A^e_j\chi_1^f \chi_1^b -\frac12 \chi^{ef}
(\partial_j\chi_1^b )
\nonumber\\
&&\qquad+\frac12 (\partial_j\chi^{ef})\chi_1^b
-\frac16(\partial_j\chi_1^e )\chi_1^f \chi_1^b \Big.\Big)\,.
\end{eqnarray}

To now calculate the potential between such charges, we take a quark
and an antiquark, both dressed according to Eq.~\ref{gthree},
average over colours and study the expectation value of the QCD
Hamiltonian. The potential is given by the dependence of the energy
on the quark separation, $r:=\vert \boldsymbol y - \boldsymbol
y'\vert$. The lowest order contribution from either charge is of
order $g$ and so, to calculate the potential at order $g^4$ between
two charges, we only need to expand the two dressings up to order
$g^3$. The potential is therefore
\begin{equation}\label{energy}
  V(r)=
  \frac1{2N^2} \int\!d^3x
  \bra{\bar\psi(y)h(y)h^{-1}(y')\psi(y')}
({E^a_i}^2(x)+{B^a_i}^2(x))
\ket{\bar\psi(y)h(y)h^{-1}(y')\psi(y')}
  \,,
\end{equation}
which implies
\begin{equation}\label{potential}
  V(r)=
  -\frac1N\textrm{tr}\int\!d^3x \bra0 [E^a_i(x),h^{-1}(y)]h(y)
  [E^a_i(x),h^{-1}(y')]h(y') \ket0
  \,,
\end{equation}
where the trace is over colour and we have used the fact that
$B_i^a$ commutes with the minimal dressing.

It follows at leading order from  (\ref{potential}) that
\begin{equation}\label{gtwo}
  V(r)=
  -\frac{g^2}{N}\textrm{tr}\int\!d^3x \bra0 [E^a_i(x),
  \chi^d_1(y)]T^dT^b
  [E^a_i(x),\chi^b_1(y')] \ket0
  \,.
\end{equation}
Inserting the fundamental equal time commutator,
$[E_i^a(x),A_j^b(y)]=i\delta^{ab}
\delta(\boldsymbol{x}-\boldsymbol{y})$, into this last equation
gives at leading order
\begin{equation}\label{gtworesult}
  V(r)=
  -\frac{g^2  C_F}{4\pi r}
  \,.
\end{equation}
We recognise the Coulombic inter-quark
potential~\cite{Fischler:1977yf, Appelquist:1977tw}.

In general, and especially at higher orders, it is simpler to work
in momentum space. Integral representations based upon the identity
\begin{equation}\label{mom}
  \frac1{(\xb^2)^a}=
\frac{4^{\frac d2-a}\pi^{\frac d2}\Gamma(\frac d2-a)}{\Gamma(a)}
\intdq\frac1{(\qb^2)^{\frac d2-a}}\ee^{i\qb\cdot\xb}
  \,,
\end{equation}
make the calculations much easier. In this way the  contribution
to the quark potential from the  minimal dressing at ${\cal
O}(g^4)$ has previously~\cite{Lavelle:1998dv} been shown to be
\begin{equation}\label{Vtwiddle}
 -3g^4C_FC_A\frac{k_l k_m}{\kb^4}
\intp \frac{i\tilde{D}_{lm}^{TT}(\pb)}{(\kb-\pb)^2}\,,
\end{equation}
where $d$ is the number of spatial dimensions ($d=3-2\epsilon$) and
the tree level equal time gluon propagator in momentum space is
given by
\begin{equation}\label{Dtwiddle}
i\tilde{D}_{lm}(\pb)=\int d^dx\,
iD_{lm}(0,\xb)e^{-i\pb\cdot\xb}\,.
\end{equation}
The superscript $T$ in~(\ref{Vtwiddle}) signifies projection upon
the transverse components, $k_i A^T_i=0$. This shows the gauge
invariance of (\ref{Vtwiddle}) and it is straightforward, if
tedious, to show that the longitudinal, gauge dependent $A_i^L$
fields cancel in this result. At order $g^4$ this corresponds to
inserting the free transverse projected, equal time propagator
\begin{equation}\label{freeprop}
\langle A_j^T(w)A_k^T(z)\rangle= \frac1{2\pi^2}\frac{(z-w)_j
(z-w)_k}{{\mod{\boldsymbol z-\boldsymbol w}}^4} \,,
\end{equation}
which yields the dominant part of the bare  potential corresponding
to anti-screening:
\begin{equation}\label{Vg4x}
\tilde V_{\mathrm{min}}^0(\kb^2)=-4\pi
C_F\frac{\alpha_0}{\kb^2}\left\{
1-\frac{\alpha_0}{\pi}C_A\left[\log\left(\frac{\kb^2}{\nu^2}\right)+2\log(2)-\frac73
-\frac1{\epsilon}\right] \right\}\,.
\end{equation}
This should be contrasted with the  bare potential (\ref{bareV}).
The difference between the divergences in these two results is due
to screening. The equations clearly show that gluons screen as well
as anti-screen. The screening effect is due to transverse, gauge
invariant glue from the additional dressing. The relative weighting
of gluonic anti-screening to screening by glue is 12 to 1. (There is
no anti-screening contribution from the matter fields at this
order.)

These effects have also been calculated~\cite{Bagan:2000nc} in
$2+1$ dimensions where it was seen that the relative weighting of
anti-screening and screening in the potential is the same within
1\%.

\subsection*{The Leading in $n_f$ Potential at Order $g^6$}

\noindent At the next order in the coupling there are
contributions from gluons and from light quarks. Here we will
calculate the quark contribution, i.e., the $n_f$ dependent terms,
to the minimal dressing.  As is well known, quarks produce at next
to leading order a screening of (electric and) colour charges and
we have seen above that, at order $g^4$, there are no
contributions from quarks to the minimal anti-screening potential.
However, we will now show that at next to next to leading order
quarks also produce an anti-screening effect. This contribution is
needed to ensure gauge invariance at higher orders. It occurs
through the one loop, fermionic correction to the gluon propagator
in (\ref{Vtwiddle}).

In addition to (\ref{Vtwiddle}) there are other contributions to
the minimal potential at order~$g^6$. They arise by higher order
expansions of the dressings and will involve Green's functions
such as $g^5\bra{0}AAA\ket{0}$ and $g^6\bra{0}AAAA\ket{0}$. These
Green's functions will only depend on the $n_f$ light fermions
through loops and it is easy to see that they will first introduce
quark contributions beyond order $g^6$ in the coupling. We
conclude that the first quark contribution to the anti-screening
potential comes from (\ref{Vtwiddle}) alone.

It should also be noted that although the QCD two point function
$\bra0A^TA^T\ket{0}$ in (\ref{Vtwiddle}) is not generally gauge
invariant at higher orders (see Appendix A
of~\cite{Lavelle:1997ty}), at one loop its $n_f$ dependent part is
indeed gauge invariant. At order $g^2$ we have the well known $n_f$
dependent term from the one loop contribution to the gluon
polarisation
\begin{equation}\label{pol}
\Pi(p)=\frac{g^2n_f}{(4\pi)^{\frac D2}} \frac{D-2}{D-1}
(-p^2)^{\frac D2-2} \frac{\Gamma(2-\frac D2) \Gamma^2(\frac D2-1)}
{\Gamma(D-2)}\,,
\end{equation}
where we skip the obvious transverse projection tensor. This
enters the one loop propagator via the contribution, $iDi\Pi iD$,
which implies
\begin{equation}
\int d^Dx\bra0 \mathrm{T}(A_i(x)A_j(0)\ket0e^{-ip\cdot x} =
-\frac{i}{(p_0^2-\pb^2)^2} [p_i p_j+\delta_{ij}(p_0^2-\pb^2)]
\Pi(p_0^2-\pb^2)\,.
\end{equation}
The one loop equal time propagator in momentum space,
$i\tilde{D}_{ij}$ is now defined to be
\begin{equation}
i\tilde{D}_{ij}(\pb) =
-\int^\infty_{-\infty}\frac{dp_0}{2\pi}\frac{i}{(p_0^2-\pb^2)^2}
[p_i p_j+\delta_{ij}(p_0^2-\pb^2)] \Pi(p_0^2-\pb^2)\,.
\end{equation}
Projecting onto the transverse components (which are gauge
invariant at this order in $g$) via $\delta_{il}-{p_ip_l}/\pb^2$
and $\delta_{jm}-{p_jp_m}/\pb^2$ gives
\begin{equation}
i\tilde{D}_{lm}^{TT}(\pb) = -\left( \delta_{lm}-\frac{p_l
p_m}{\pb^2}\right)
\int^\infty_{-\infty}\frac{dp_0}{2\pi}\frac{i}{(p_0^2-\pb^2)}
\Pi(p_0^2-\pb^2)\,.
\end{equation}
Inserting (\ref{pol}) yields
\begin{equation}\label{ref}
i\tilde{D}_{lm}^{TT}(\pb) = -g^2n_f\left( \delta_{lm}-\frac{p_l
p_m}{\pb^2}\right) \frac{1}{(\pb^2)^{2-\frac d2}}
\frac1{2^{2+d}\pi^{\frac d2}}
\frac{\Gamma(3-d)\Gamma(\frac{1+d}2)}{\Gamma(\frac{5-d}2)\Gamma(\frac{2+d}2)}
\,.
\end{equation}

To calculate the $n_f$ dependent part of the potential, we now
insert this into (\ref{Vtwiddle}). It is helpful, though, to
rewrite the resulting expression via
\begin{equation}\label{simpler}
1-\frac{(\kb\cdot \pb)^2}{\kb^2\pb^2} = \frac14 \left(
2-\frac{\kb^2}{\pb^2}-\frac{\pb^2}{\kb^2}\right)+\dots \,,
\end{equation}
where we have \emph{dropped} terms that only contribute massless
tadpoles in the subsequent integral and will hence vanish in
dimensional regularisation. This leads to the order $g^6$
contribution to the potential~(\ref{Vg4x})
\begin{equation}\label{Vnf}
\frac{g^6n_fC_FC_A}{\kb^2}\frac3{2^{4+d}\pi^{\frac{d}2}}
\frac{\Gamma(3-d)\Gamma(\frac{1+d}2)}{\Gamma(\frac{5-d}2)\Gamma(\frac{2+d}2)}
\intp\frac1{(\pb^2)^{2-\frac d2}(\kb-\pb)^2} \left(
2-\frac{\kb^2}{\pb^2}-\frac{\pb^2}{\kb^2}
 \right)
\,.
\end{equation}
The divergent part of this contribution, in terms of the bare
coupling $\alpha_0$, is:
\begin{equation}\label{Vnfa}
 \frac{\alpha_0^3n_fC_FC_A}{3\kb^2\pi}
\left\{ \frac1{\epsilon^2}+\frac1\epsilon\left[ \frac{14}3-2\gamma_E
-2\log\left(\frac{\kb^2}{4\pi\nu^2}\right) -4\log(2)
 \right]
 \right\}
\,.
\end{equation}
Note that these divergences are ultra-violet singularities as can be
seen from power counting in~(\ref{Vtwiddle}) and~(\ref{ref}). The
leading singularity here is local, but the sub-leading divergences
include the $\frac1\epsilon\log(\kb^2)$ term which is a non-local
divergence. The immediate question is can renormalisation deal with
this infinity?

\subsection*{Renormalising the Minimal Potential}

The minimal part of the inter-quark potential has been previously
calculated at order $g^4$ in both four and three dimensions. The
result in four dimensions (\ref{Vg4x})  may be renormalised by
using \emph{minimal} charge renormalisation where we define the
minimally renormalised coupling, $\alpha'$, through
$\alpha_0=Z^{\mathrm{min}}_{\alpha'}\alpha'$ where
\begin{equation}\label{Zmin}
Z_{\alpha'}^{\mathrm{min}}= 1-\frac{\alpha'}{\pi}C_A
\frac1{\epsilon}\,.
\end{equation}
This anti-screening renormalisation is defined so that
\begin{eqnarray*}
\tilde V_{\mathrm{min}}(\kb^2) &=&-4\pi
C_F\frac{\alpha'}{\kb^2}\left[ 1-\frac{\alpha'}{\pi}C_A
\frac1{\epsilon}
\right] \\
&&\qquad\times \left\{
1-\frac{\alpha'}{\pi}C_A\left[\log\left(\frac{\kb^2}{\nu^2}\right)
-\frac1{\epsilon}\right] \right\}\,,
\end{eqnarray*}
is finite at this order in the minimal coupling:
\begin{equation}\label{Vminfinite}
\tilde V_{\mathrm{min}}(\kb^2) =-4\pi C_F\frac{\alpha'}{\kb^2}
\left\{
1-\frac{\alpha'}{\pi}C_A\log\left(\frac{\kb^2}{\nu^2}\right)
\right\}+{\cal O}(\alpha'^3)\,.
\end{equation}
This is a very direct way to extract the minimal, anti-screening
beta function which has also been observed in very different
ways~\cite{Drell:1981gu}. This minimal coupling clarifies the
nature and importance of anti-screening in non-abelian gauge
theories.

It is, however, very simple to show that the \lq anti-screening
coupling\rq\ cannot be used at next order. We have seen that there
is a non-local, $n_f$ dependent divergence in the minimal
potential at order $g^6$ and this is the \emph{only} $n_f$
dependence in the minimal potential (\ref{Vnfa}) up to this order.
There is a logarithm  at order $\alpha_0^2$ which might help
produce non-local divergences at $\alpha_0^2$ but it is not $n_f$
dependent  and in the leading anti-screening charge
renormalisation (\ref{Zmin}) there is no $n_f$ dependence either.
Any $n_f$ dependence in $Z_{\alpha'}^{\mathrm{min}}$ at order
${\alpha'}^3$ would, of course, be local and not introduce any
logs into the potential at order ${\alpha'}^3$. Thus
\emph{nothing} can cancel the non-local divergence at order
${\alpha'}^3$ in this approach. We are forced to conclude that the
anti-screening or minimal charge renormalisation of the minimal
potential breaks down beyond leading order. It is, in other words,
impossible to define a coupling renormalisation scheme that
renormalises the minimal, anti-screening potential. We cannot,
beyond lowest order, speak of screening and anti-screening
structures in the beta function.

It is, however, not necessary to use the anti-screening coupling in
the minimal potential. Instead one can use full coupling
renormalisation (\ref{Zfull}) plus an additional multiplicative
renormalisation of the minimal potential in (\ref{Vg4x}):
\begin{equation}\label{Vren}
\tilde V_{\mathrm{min}}=\nu^{-2\epsilon}Z_V\tilde
V^0_{\mathrm{min}}(\kb^2)
\end{equation}
where we write
\begin{equation}\label{Vrenstruc}
Z_V=1+\delta^1_V\frac{\alpha}\pi+\delta^2_V\left(\frac{\alpha}\pi\right)^2+\dots
\end{equation}
The minimal potential is easily seen to be finite at this order if
\begin{equation}\label{Vrenstruc1}
\delta^1_V=-\left(\frac1{12}C_A+\frac{n_f}6\right)\frac1{\epsilon}\,.
\end{equation}
This corresponds to
\begin{equation}\label{Vrenstruc2}
\tilde V_{\mathrm{min}} =-4\pi C_F\frac{\alpha}{\kb^2} \left\{
1-\frac{\alpha}{\pi}C_A\log\left(\frac{\kb^2}{\nu^2}\right)
\right\}+{\cal O}(\alpha^3)\,.
\end{equation}
Our interpretation of this additional factor, $Z_V$,  is that it
is a renormalisation of the additional potential energy between
excited, minimally dressed charges compared to the true ground
state of the fully dressed system, i.e., with screening effects
included.

We will now show that this second approach may still be used at
the next order of perturbation theory, i.e., the minimal potential
is indeed renormalised by the full coupling (\ref{Zfull}) and the
potential renormalisation of (\ref{Vren}) and
 (\ref{Vrenstruc}).  At  order $\alpha^3$ scheme
dependence appears and we use the $\overline{\mathrm{MS}}$ scheme.
We require the standard two loop coupling renormalisation
\begin{equation}\label{zeds}
Z_\alpha=1+z^1_\alpha\frac\alpha\pi+z^2_\alpha\left(\frac\alpha\pi\right)^2
\,,
\end{equation}
where
\begin{equation}\label{zedtwo}
z^2_\alpha= \frac1{\epsilon^2}\left(\frac{11C_A}{12}-
\frac{n_f}{6}\right)^2 -\frac1\epsilon\left( \frac{17C_A^2}{48} -
\frac{C_Fn_f}{16} -\frac{5C_An_f}{48} \right)
 \,.
\end{equation}
We now define
\begin{equation}\label{deltas}
\delta_V^2=\delta_V^{2a}\frac1{\epsilon^2}+
\delta_V^{2b}\frac1{\epsilon}\,.
\end{equation}
At order $n_f\alpha^3$ in the potential, we first consider the
$1/\epsilon^2$ terms. Inserting all the above renormalisation
constants and demanding the cancellation of $1/\epsilon^2$ terms
leads to
\begin{equation}\label{delta2aresult}
\delta_V^{2a}= C_A\left(\frac2{3}C_A+\frac{n_f}{12}\right) \,.
\end{equation}
(Note that the $n_f$ independent term must be corrected by gluonic
anti-screening effects which we neglect.)

Inserting this into the potential and demanding the vanishing of
the local $1/\epsilon$ terms yields
\begin{equation}\label{delta2bresult}
\delta_V^{2b}= -n_f\left(\frac5{48}C_A+\frac{C_F}{16}\right) \,,
\end{equation}
plus various $n_f$ independent terms from the purely gluonic
contributions to the anti-screening potential.

Having now fixed the renormalisation constant, it is very
satisfying to see that the non-local divergences in (\ref{Vnfa})
are cancelled in this scheme. At order $\alpha^3$ there are three
such non-local terms: they are generated by $n_f$ dependent local
divergences in the renormalisation constants multiplying the
logarithm in the one loop potential (\ref{Vg4x}). One is from the
$n_f$ part of $Z_V$:
\begin{equation}\label{pos}
-\frac{4C_F}{\kb^2}\frac{\alpha^3}\pi\frac{n_f}6
\log\left(\frac{\kb^2}{\nu^2}\right)\frac1{\epsilon}\,,
\end{equation}
while there are two further $n_f$ dependent contributions from the
coupling constant renormalisation (\ref{Zfull}) since $\alpha_0$
occurs twice in (\ref{Vg4x}). Each of these yields
\begin{equation}\label{pos1}
-\frac{2C_F}{\kb^2}\frac{\alpha^3}\pi\frac{n_f}6
\log\left(\frac{\kb^2}{\nu^2}\right)\frac1{\epsilon}\,,
\end{equation}
and adding all three of these terms together we see that the $n_f$
dependent, non-local divergences in $V_{\mathrm{min}}$ at order
$\alpha^3$ indeed cancel. We stress that this cancellation is a
stringent test of the method since there was no freedom in the
calculation. We conclude that this renormalisation programme can
be carried through.

Our final result for the renormalised, anti-screening potential is
\begin{equation}\label{potdef0}
\tilde V_\mathrm{min}(\kb^2)=-\frac{4\pi\alpha
C_F}{\kb^2}+\alpha^2\tilde V^2_\mathrm{min}(\kb^2) +\alpha^3\tilde
V^3_\mathrm{min}(\kb^2)+\dots\,,
\end{equation}
where
\begin{equation}\label{potres1}
\tilde V^2_\mathrm{min}(\kb^2)= \frac{4C_AC_F}{3\kb^2}\left(
-7+6\ln(2)+3\ln\left(\frac{\kb^2}{\nub^2}\right) \right)\,,
\end{equation}
and the $n_f$ dependent terms
\begin{eqnarray}\label{potres2}
\tilde V^3_\mathrm{min}(\kb^2)&=& \frac{C_AC_Fn_f}{27\pi\kb^2}
\Bigg(\Bigg. 125-3\pi^2
+12\ln(2)[-7+3\ln(2)] \\
&&\left.\quad+3\ln\left(\frac{\kb^2}{\nub^2}\right)
\left[-14+12\ln(2)+3\ln\left(\frac{\kb^2}{\nub^2}\right)\right]
\right)\,,
\end{eqnarray}
where $\nub^2=4\pi \nu^2e^{-\gamma_E}\,$.
\subsection*{Conclusions}

\noindent We have seen that the decomposition of the beta function
into screening and anti-screening structures breaks down beyond
one loop. This we saw by calculating the light quark contributions
to the anti-screening potential: non-local divergences arose in
fermion loops in the minimal potential at order $g^6$ which are
not cancelled by an anti-screening beta function. This is due to
anti-screening effects from light fermions which are necessary
consequences of a gauge invariant construction of charges.

However, we have seen that it is possible to renormalise this
potential via full charge renormalisation plus a multiplicative
renormalisation of the potential. This renormalisation provided a
stringent test of the method. It is to be understood as a
renormalisation of the additional energy due to the neglect of
screening interactions in a minimally dressed construction of
charges.

The results presented here suggest that a decomposition of the
potential into a minimal, anti-screening part plus an additional
screening structure is indeed possible. Further studies of this
decomposition may  help
 to clarify the structure of the forces between heavy
quarks.

\medskip

\no\textbf{Acknowledgements:} We thank Tom Heinzl, Tom Steele
 and Shogo Tanimura for related discussions. ML thanks the
 Universitat Auton\`{o}ma de Barcelona for their hospitality.


\end{document}